\documentclass[aps,prl,twocolumn,superscriptaddress,showpacs]{revtex4}
\usepackage{graphicx}

\begin{document}

\title{First-principles study of phenyl ethylene oligomers as current-switch}
\author{F. Jiang}
\affiliation{Department of Physics, Fudan University, Shanghai
200433, People's Republic of China}
\author{Y.X. Zhou}
\affiliation{Department of Physics, Fudan University, Shanghai
200433, People's Republic of China}
\author{H. Chen }
\email[Corresponding author. Email: ] {haochen@fudan.edu.cn}
\affiliation{Department of Physics, Fudan University, Shanghai
200433, People's Republic of China}
\author{R. Note}
\affiliation{Institute for Materials Research, Tohoku University,
Sendai 980-8577, Japan}
\author{H. Mizuseki} \affiliation{Institute for Materials Research, Tohoku
University, Sendai 980-8577, Japan}
\author{Y. Kawazoe}
\affiliation{Institute for Materials Research, Tohoku University,
Sendai 980-8577, Japan}

\date{\today}

\begin{abstract}
 We use a self-consistent method to study the distinct current-switch of
$2^{'}$-amino-$4$-ethynylphenyl-$4'$-ethynylphenyl-$5'$-nitro-$1$-benzenethiol,
from the first-principles calculations. The numerical results are
in accord with the early experiment [Reed et al., Sci. Am.
\textbf{282}, 86 (2000)]. To further investigate the transport
mechanism, we calculate the switching behavior of p-terphenyl with
the rotations of the middle ring as well. We also study the effect
of hydrogen atom substituting one ending sulfur atom on the
transport and find that the asymmetry of I-V curves appears and
the switch effect still lies in both the positive and negative bias
range.
\end{abstract}

\maketitle
 As Moore's law predicted, the density of transistors
 on integrated circuits doubles approximately every 18 months,
 and the traditional semiconductor devices will approach to their limit.
 In this century, the development of electronics may enter a new
 era of molecular devices. Compared with the traditional solid electronic devices,
 the molecular devices have many
 advantages. The molecular chips will be smaller than the silicon chips in three orders
 and their operation speed will increase evidently without increases
 in price. The concept of the organic molecular electronic devices was first proposed over a
 quarter century ago by Aviram and Ratner \cite{1}. However, the molecular devices
 did not attract much attention until
 the measurement of the electronic conduction through the single phenyl dithiol (PDT)
 by Reed et al. in 1997 \cite{2}.
 From then on, the search for new active molecular devices becomes a worldwide
 effort \cite{3,4,5,6,7,8,9,10,11,12,13,14,15}.
 Molecules are intrinsically different from traditional devices, and one expects that some of their transport properties
 would be unique. An attractive feature of organic molecules for application to electronics is the possibility
 to control electronic transport by using electrical or photic signal. Several experimental groups
 reported some molecular switch devices \cite{16,17,18},
 which are mainly included into two categories. One is the
 electronic switch devices which control the conversion between state
 \textquoteleft\textquoteleft$0$\textquoteright\textquoteright
 \ and state \textquoteleft\textquoteleft$1$\textquoteright\textquoteright
 \ by the reversible field or the current pulse. The other is the
  photic switch devices which achieve the open or close states of
  current by the laser field. Reed et al. \cite{19}
  used a molecular switch consisting of three aromatic phenyl rings in series.
  The two hydrogen atoms of the middle ring are substituted
  by acceptor group $\mbox{NO}_{2}$ and donor group
  $\mbox{NH}_{2}$, while the whole molecule is chemisorbed onto the contact surfaces of
  gold leads. By adjusting the gate voltage, the current can be
  controlled due to rotation of the middle ring with respect to the side rings.
  The corresponding theoretical works investigated the
  similar switch system \cite{20,21,22}.
  However, in these works, the macroscopical electrodes
  were not treated on an equal footing with the molecular devices,
  and the whole open system, lead-molecule-lead, were not treated self-consistently
  with the charge
  effect included. In the quantum system, the charge
  effect is crucial to the transport property. The rigorous treatment
  of the molecular device calls for
  combining the theory of quantum transport with the first-principles calculations
  of the electronic structure. In this article,
  we use density functional theory (DFT) and nonequilibrium Green's
  function to study the current-switch behavior of the phenyl ethylene oligomers.

    For the lead-molecule-lead system, coupling between the molecule and the
  leads plays a crucial role in quantum
  transport. Being computationally tractable, the whole system is partitioned
  into the molecule part and the
lead part, so that these two parts are dealt with separately. The
nonequilibrium Green's function theory provides a powerful method
to give a full description of transport phenomena. The molecular
Green's function is expressed as follows
\begin{equation}G^{R}_{M}=(E^{+}S_{M}-F_{M}-\Sigma^{R}_{1}-\Sigma^{R}_{2})^{-1},
\end{equation}
where $G^{R}_{M}$, $S_{M}$, $F_{M}$ are the retarded Green's
function, overlap matrix and Fock matrix of the molecule part
respectively. $\Sigma^{R}_{1}$ ($\Sigma^{R}_{2}$), the retarded
self-energy of the left (right) electrode, is calculated from the
surface Green's function (SGF) $g^{R}_{1}$ ($g^{R}_{2}$)
\begin{equation}\Sigma^{R}_{i}=(E^{+}S_{Mi}-F_{Mi})g^{R}_{i}(E^{+}S_{iM}-F_{iM}),
\end{equation}
with $i=1,2$. The coupling matrices $S_{Mi}$ and $F_{M,i}$ are
extracted from the DFT calculation for the extended molecule
(molecule with 3 Au atoms on each side).

The density matrix of the open system is the essential function of
the whole self-consistent scheme. It can be achieved by the
Keldysh Green's function
\begin{equation}\rho=\int^{\infty}_{-\infty} dE[-iG^{<}(E)/2\pi],\end{equation}
\begin{equation}-iG^{<}=G^{R}_{M}(f_{1}\Gamma_{1}+f_{2}\Gamma_{2})G^{A}_{M},
\end{equation}
with the advanced Green's function $G^{A}=(G^{R})^{\dagger}$, the
broadening function of the left (right) lead $\Gamma_{1}$
($\Gamma_{2}$). The Fermi distribution function of the left
(right) lead $f_{1}$ ($f_{2}$) is expressed
$f_{i}(E)=1/(e^{(E-\mu_{i})/kT}+1)$ with
$\mu_{1}=\mbox{$E_{f}$}-\frac{1}{2}eV$,
$\mu_{2}=\mbox{$E_{f}$}+\frac{1}{2}eV$, we regulate Z axis
pointing from the left to the right, so $V>0$ ($V<0$) denotes the
longitudinal electrical field direction along the positive
direction of Z axis (along the negative direction of Z axis). But,
this expression is only right for the case that the molecule is
symmetrical to the two contacts. If not symmetrical, the voltage
drop is also not symmetrical to the two interfaces. Fermi level of
the bulk Au is $E_{f}$. In our work, $E_{f}$ is -5.1 eV which is
adjusted around its work function (5.31 eV). Since the relative
dielectric constant of the molecule is much larger than 1, the
voltage drops across the interface between gold atoms and the
sulfur atom, while the electronic potential is almost flat. So,
the potential zero point can be set at the center of the molecular
device.

After obtaining the density matrix self-consistently, we
calculate the transmission function $T(E,V)$ of coherent transport
\begin{equation}T(E,V)=\mbox{Tr}(\Gamma_{1}G^{R}\Gamma_{2}G^{A}).\end{equation}
In order to achieve this goal, we extended the inner loop in the
standard quantum chemistry software Gaussian03 \cite{23} to the
loop containing the lead-molecule-lead open system under bias. The
self-consistent procedure starts from a guess for the density
matrix of the open system, which may be obtained from the converged
density matrix of Gaussian03 calculation for the isolated
molecule. We feedback the density matrix to the Gaussian's
main program as a subroutine to obtain the new density matrix. The iterations
continue until the density matrix converges to the acceptable
accuracy. Then the density matrix is used to evaluate the terminal
current and density of states (DOS) of the open system
\cite{24,25}. In the calculations, we adopt DFT with B3PW91
exchange-correlation potential and LANL2DZ basis to evaluate the
electronic structure and the Fock matrix. The basis set associates
with the effective core potential (ECP), which are specially
suited for the fifth-row (Cs-Au) elements with including the
Darwin relativistic effect.

  In Reed's experiment \cite{19}, the middle ring of the molecule
  $2^{'}$-amino-$4$-ethynylphenyl-$4'$-ethynylphenyl-$5'$-nitro-
  $1$-benzenethiol can rotate with respect to the side benzene rings
  under control of the gate voltage.
  The dipole composed of the acceptor NO$_2$ and the
  donor NH$_2$, which are attached to the middle benzene, is driven by the external field.
  Our DFT calculation at LANL2DZ level achieves the optimized molecular structure,
  where three aromatic phenyl rings are almost in the same plane,
  the carbon bond length of the alkyne
  in the middle of two rings is 1.24 \AA, the angle of O-N-O is $128^{\circ}$
  and the angle of H-N-H is $118^{\circ}$.
  For the sulfur atoms chemisorbed onto the gold leads, there are two
  kinds of the geometric configurations. The sulfur atom sits
  directly on the top position of the surface gold atom or the hollow position
  of three nearest-neighbor surface gold atoms. As a conventionally accepted
  picture the latter is adopted in this article. The perpendicular distance between
  the sulfur atom and the Au FCC (111) surface plane is 2.0 \AA,
  an usually acceptable distance. Due to the precalculation, the
  temperature effect is not distinct for the short molecule, so,
  we assume zero temperature in our calculation for simplicity. It is emphasized that
  since our system is a closed shell system (the number of $\alpha$
  electrons is equal to that of $\beta$ electrons), we adopt
  restricted calculation which ensures that
  $\rho^{\alpha}=\rho^{\beta}$ in order to spare calculation time.
  We investigate the molecules under three configurations corresponding
  to the rotation angle $\alpha=0^{\circ}$, $30^{\circ}$, and $60^{\circ}$ respectively.
  DOS and T corresponding to the different values of $\alpha$ are given in Fig. 1,
  which accounts for the mechanism of the molecular switch.
Both HOMO and LUMO are shifted and broadened with the molecule attached
by the metallic leads,
so that $E_{f}$ is almost in the middle of HOMO-LUMO gap, a little closer to
HOMO, which means electrons will be responsible for the initial
rise of the current. The electron incident from the left lead with
the energy meeting the one of molecular states will enter the
resonant channel. The states with
 small DOS in the HOMO-LUMO gap is called as the metal-induced gap states (MIGS)
 arising from the hybridization of gold
 surface states and molecular HOMO, LUMO states. They offer some electrons
 (about 0.32 electrons) to the contacts,
which causes the
  molecule to be a system with positive charges. When bias is
  applied, the system in the nonequilibrium state adjusts its energy levels
  by the charge effect to prevent the loss of electrons.

  The transport characteristics of the molecule-lead system
  with the rotated configurations are obtained from the DOS
  curves. The singular points of the Green's function,
  which are obtained from equation
  $(F+\Sigma_1+\Sigma_2)C=SC\lambda$,
  show the position of the levels of the open system.
  In the case of $\alpha=0^{\circ}$, HOMO is -6.53 eV, LUMO is -3.60 eV, and the
  HOMO-LUMO gap is 2.93 eV. In the case of $30^{\circ}$, HOMO is -6.54 eV,
  LUMO is -3.56 eV, and the HOMO-LUMO gap is 2.98 eV.
  And for $60^{\circ}$,
  HOMO, LUMO and the HOMO-LUMO gap are -6.73 eV, -3.51 eV, and 3.22 eV, respectively.
  With the increase in the rotation angle, HOMO and
  LUMO of the open system keep away from $E_{f}$ step by step,
  and the transmission function value at the Fermi level
  gets smaller. The transmission functions are obtained for
  the three configurations,
  $T(E_{f})=0.19\times10^{-2}$, $0.14\times10^{-2}$,
$0.34\times10^{-3}$ corresponding to $0^{\circ}$, $30^{\circ}$ and
$60^{\circ}$. The big rotation angle (e.g. $60^{\circ}$) results
in the heavy decrease in the current (the $90^{\circ}$ rotation
causes the switch into the off state, not shown in the figure)
which demonstrates the switchable function of the molecular
device.
  With the molecular device attached by the gold leads,
  the energy levels are broadened and descend to
  keep the system electrical-neutral. The electron number
  inside the device decreases from 186 (in the isolated molecule)
  to 185.6. The basic reason is illustrated by the inset of Fig.
  2. The inset shows the energy levels of the isolated molecule
  corresponding to the three cases. LUMO of the isolated molecule is
  below the Fermi level, and LUMO+1 is above the Fermi level. With
  the increase of the rotation angle, LUMO-LUMO+1 gap gets bigger,
  leading to the same tendency shown in Fig. 1.
  With leads attached, the charge effect makes the energy levels
  of the isolated molecule descend and broadened. The LUMO-LUMO+1 gap
  evolves into the HOMO-LUMO gap of the open system and
  determines the transport capacity. In Fig. 2 the I-V curves are steep
  for $\alpha\le 30^{\circ}$,
  since both HOMO and LUMO
  are responsible for the rising of current.
  The negative differential resistance (NDR) appears with the
  added electrons, when bias gets across -3.6 V. From $30^{\circ}$ to $60^{\circ}$,
  the switch effect due to
  the rotation is more apparent
  than the case $\alpha \le 30^{\circ}$.

  \begin{figure}
\includegraphics[scale=0.45,bb=16 13 565 801]{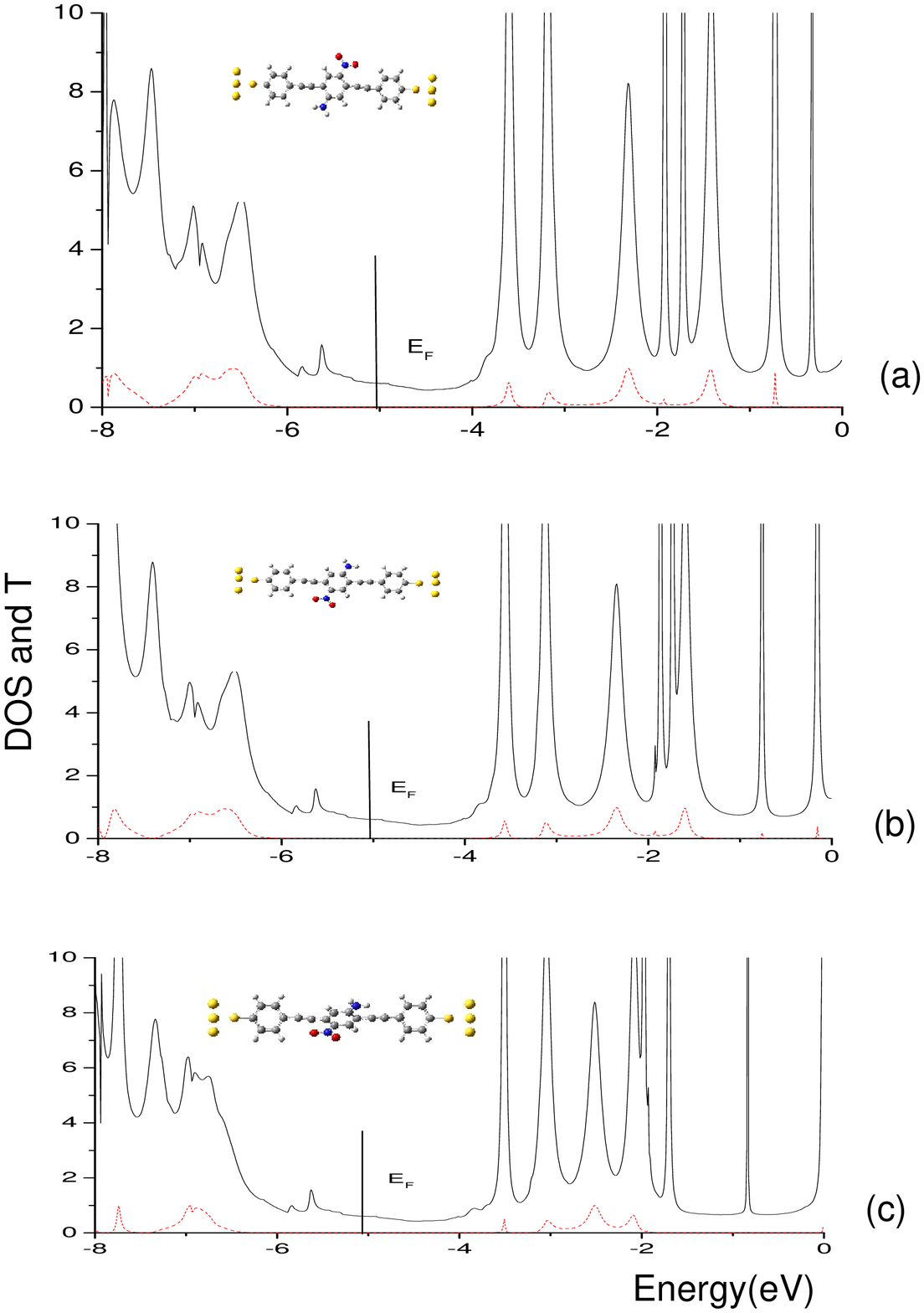}
\caption{$2^{'}$-amino-$4$-ethynylphenyl-$4'$-ethynylphenyl-$5'$-nitro-$1$-benzenethiol
is chemisorbed onto the gold leads through sulfur atoms from both
sides. DOS (solid) and T (dashed) as functions of energy in
equilibrium. The vertical line denotes the position of Fermi
level. (a) rotation angle $\alpha=0^{\circ}$, (b)
$\alpha=30^{\circ}$, (c) $\alpha=60^{\circ}$.\label{fig1}}
\end{figure}

  \begin{figure}
\includegraphics[scale=0.35,angle=-90,bb=28 42 574 748]{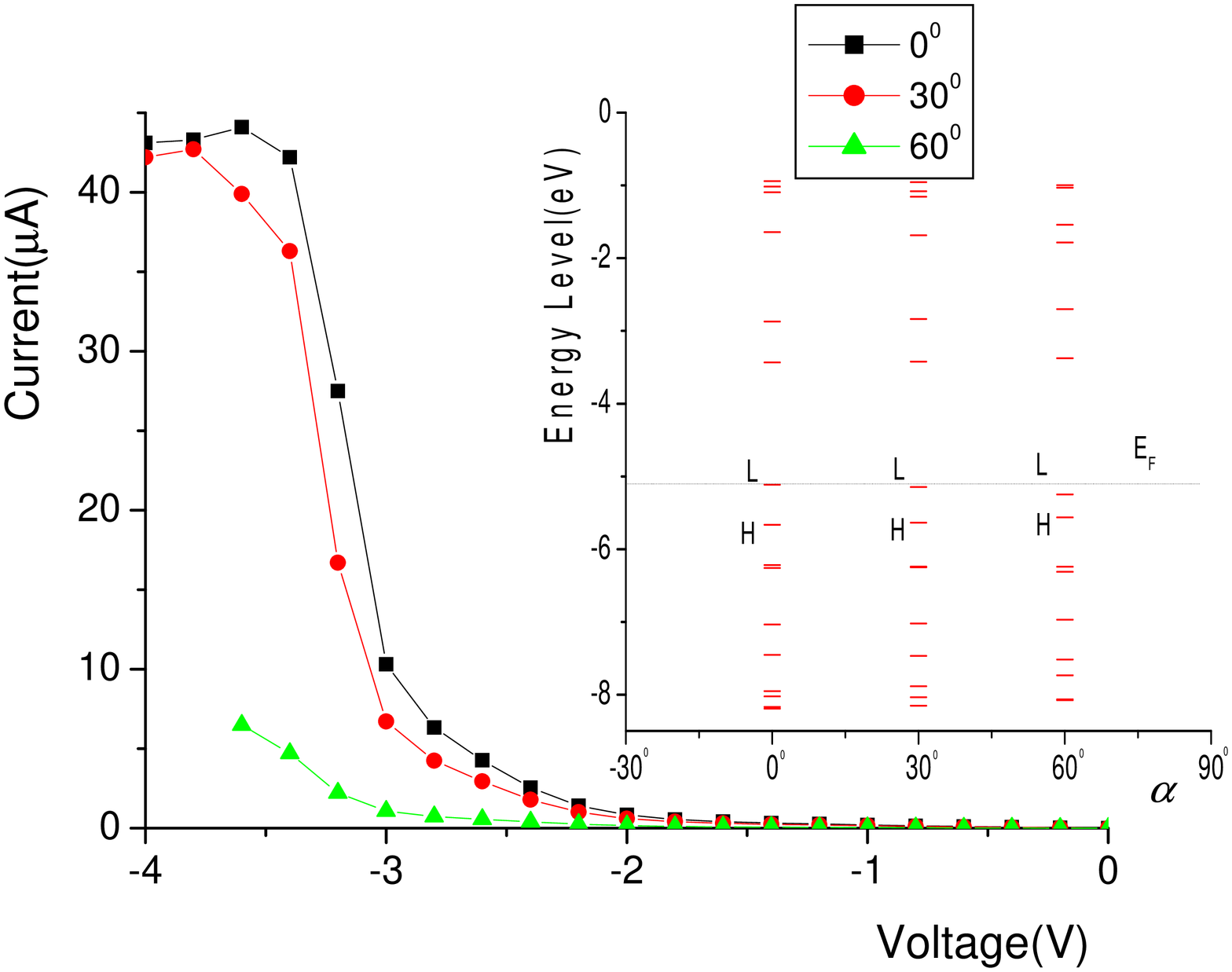}
\caption{I-V curves of
$2^{'}$-amino-$4$-ethynylphenyl-$4'$-ethynylphenyl-$5'$-nitro-$1$-benzenethiol
with gold contacts through sulfur atoms from both sides
corresponding to $\alpha=0^{\circ}$, $30^{\circ}$ and
$60^{\circ}$.\label{fig2}}
\end{figure}

  The p-terphenyl is a good sample to investigate the molecular switch properties.
  Fig. 3 gives the I-V curves of p-terphenyl with the anti-clockwise rotations of
  the middle ring with respect to the side benzenes. In the figure
  there are five current curves corresponding to rotation angle
  $\alpha=0^{\circ},$  $30^{\circ},$  $45^{\circ},$  $60^{\circ},$ and $90^{\circ}$.
  The molecular device presents an active switch function, although it is not easy
  to make it work in practice.
  The total energies of the isolated molecule corresponding to these
  angles are -712.967, -712.962, -712.952, -712.938,
  and -712.924 Hartree, respectively. The energy increases
  with an increase in the rotation angle, while the molecule with
  the zero angle is the most stable one. The figure illustrates a
  distinct switch function. From $0^{\circ}$ to $90^{\circ}$,
  the overlap between $\pi$ electron clouds
  decreases with the increase in the rotation angle, which leads the current
  reduction. For the $0^{\circ}$ case,
  the current of p-terphenyl under the bias 3.0 V is smaller than the one of single PDT,
  while it is bigger than the latter one above the bias 3.0 V.
  The abnormal phenomenon comes from the fact that HOMO of p-terphenyl is closer to $E_{f}$.
  The current has a big rise at the point where
  the voltage gets across HOMO level. The inset shows the molecular orbital pictures
  of HOMO for rotation angle $0^{\circ}$ and $90^{\circ}$.
  Here, the extended molecule (3 Au atoms on each side are
  included) is adopted to simulate p-terphenyl connected to
  gold contacts and its HOMO level has the main
  contribution to transport in our bias range, while LUMO does not participate in.
  It is obvious that the HOMO is a delocalized state for $0^{\circ}$ and it becomes
  a localized one when the rotation angle approaches to $90^{\circ}$.

  \begin{figure}
\includegraphics[scale=0.35,angle=-90,bb=13 58 576 743]{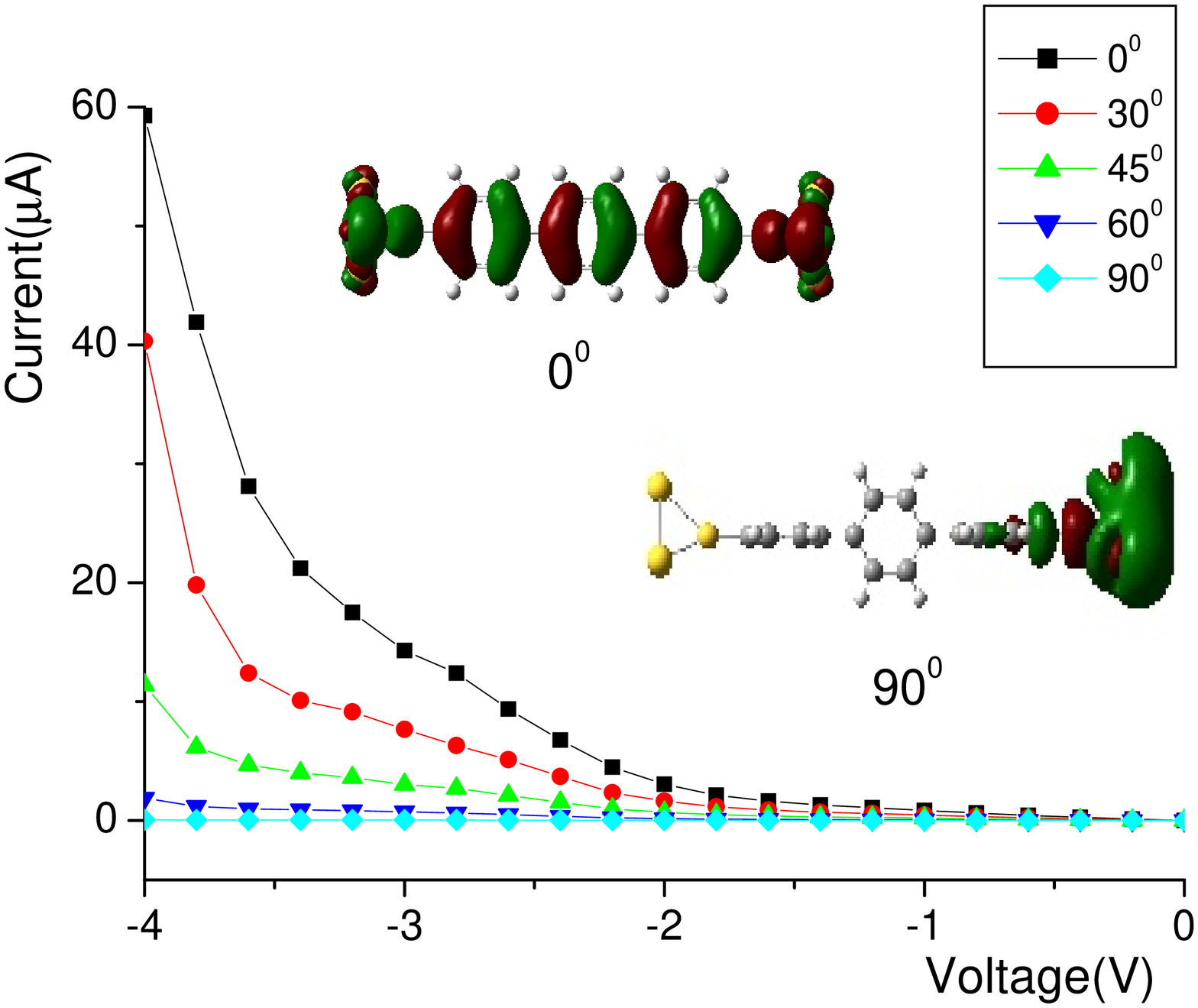}
\caption{I-V curves of extended p-terphenyl compound correspond to
the different rotation angles. The negative bias means the
electrical field is along the direction opposite to the Z
axis.\label{fig3}}
\end{figure}

  Typically the aromatic phenyl molecule is chemisorbed on the
  gold surface through a sulfur atom. J. Chen et al. \cite{26}
  once investigated the system with a hydrogen atom to
  replace the sulfur atom, and obtained a different transport
  characteristics, since the electronegativity of hydrogen atom is smaller than
  that of the sulfur
  atom, which makes the coupling between a hydrogen atom and one gold
  contact weaker than that between a sulfur atom and the other
  gold contact. The case is attractive because
  of the asymmetry of I-V curves \cite{27}.
  We add a hydrogen atom on the sulfur atom at the right end to make the whole molecule
  a closed shell system to spare the
  calculation time, without the intrinsical influence
  on I-V curve. The distance between hydrogen atom at the left end and
gold contact is a little longer than that between sulfur atom and
gold contact. But this distance is not clear. We fix this distance
at 2.15 \AA. The distance between sulfur atom and the other gold
contact is still fixed at 2.0 \AA. So, the electrostatic potential
difference V is divided between the two junctions from the
coupling distances: 52\% across the H-Au junction and 48\% across
the S-Au junction. The inset of Fig. 4 gives the energy levels of
the isolated molecule with the related HOMO-LUMO gaps: 3.15 eV,
3.23 eV and 3.43 eV for $\alpha=0^{\circ}$, $30^{\circ}$ and
$60^{\circ}$. The HOMO-LUMO gap increases with the increase in
rotation angle, which dominates the transport behavior. Fig. 5
describes DOS and T in equilibrium with the different rotation
angle. With the gold leads attached to the molecular device, there
are about 181.2 electrons in the device area. Different from the
pure S-Au contacts, only LUMO is responsible for the initial rise
of the current. With contacts, for the case of $0^{\circ}$, HOMO
is -8.03 eV, LUMO is -4.73 eV, and HOMO-LUMO gap is 3.30 eV; for
$30^{\circ}$, HOMO is -8.06 eV, LUMO is -4.68 eV, and HOMO-LUMO
gap is 3.38 eV; for $60^{\circ}$, HOMO, LUMO and HOMO-LUMO gap are
-8.08 eV, -4.62 eV, and 3.46 eV, respectively. The corresponding
I-V curves are shown in Fig. 4 for bias region -1.4 V$\le$ V $\le$
3.4 V. The distinct switch effect exists in the small negative
bias, while the electron number descends with the increase in
bias. The similar switch property occurs in the positive high bias
(above 2.0 V) with the electron number variation from 181.2 to
181.6. Because the molecule is not symmetrically connected to the
gold contacts, the negative bias is easier to get across LUMO than
the positive one and for the negative bias the charge effect does
not occur since the left chemistry potential is higher than the
right one. Then, LUMO absorbs electrons from the left side, and
emits electrons to the right side. The left coupling is weaker
than the right one, which keeps LUMO normally unoccupied and
results in the descendence of electron number, without the charge
effect. So, the current has the initial rise for the small
negative bias. For the positive bias, the left chemistry potential
is lower than that of the right one, which makes LUMO absorb
electrons from the right side and emit electrons to the left one.
The abnormally occupied LUMO due to the nonsymmetric coupling
causes the rise of electron number. So, the LUMO is kept
unoccupied by the charge effect, which results in very small
current in the small positive bias range ($V<2.0$ V). With high
bias added, the resonant transmission occurred. This process is
shown in Fig. 6. With the increase of positive bias from 1.0 V to
3.4 V, the peaks of DOS and T shift to the high energy side
continuously. At 3.4 V, the current corresponding to $0^{\circ}$
has a sudden rise since the peaks' shift makes both HOMO and LUMO
responsible for the transport at this condition. The sudden rise
makes the switch effect more apparent in comparison with the case
of $60^{\circ}$ rotation, where only LUMO is responsible for the
rise of the current. Since the distance between the hydrogen atom
and the gold contact is not clear, we also calculate the DOS and T
of the molecule in equilibrium with the H-Au distance 2.3 \AA.
Corresponding to $0^{\circ}$, $30^{\circ}$, $60^{\circ}$, the
HOMO-LUMO gaps are 3.31 eV, 3.34 eV, and 3.43 eV respectively. The
small change of H-Au distance has little effects on the positions
of peaks of DOS and T, only causes current smaller. In fact, even
this distance is fixed at 2.9 \AA, the switch effect and the
asymmetry of I-V curves still exist. Our calculation shows that
the asymmetry effect and switch effect are both stable
corresponding to the different H-Au distances. We can draw the
conclusion: for the LUMO-based molecule, the current is lower for
positive bias on the weak contact, while for the HOMO-based
molecule, the current is lower for positive bias on the strong
contact. The conclusions from the first-principles calculation
accord with the previous results obtained by the extended Huckel theory
(EHT) \cite{27}.  This asymmetry in I-V curves may be
used to provide a rectifier.

\begin{figure}
\includegraphics[scale=0.35, angle=-90, bb=19 58 562 772]{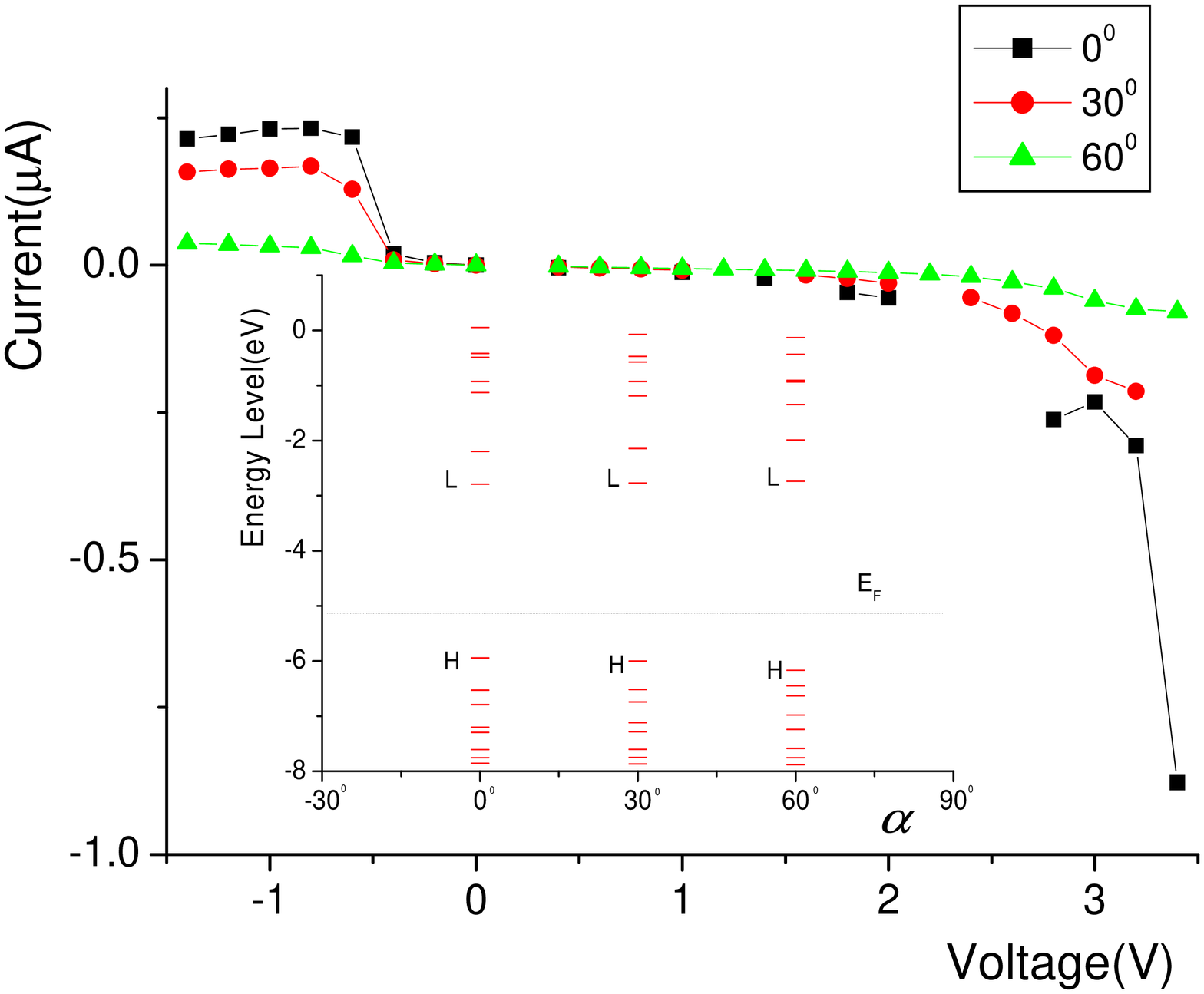}
\caption{I-V curves of
$2^{'}$-amino-$4$-ethynylphenyl-$4'$-ethynylphenyl-$5'$-nitro-$1$-benzenethiol
with gold contacts corresponding to $\alpha=0^{\circ}$,
$30^{\circ}$ and $60^{\circ}$. The atoms that chemisorbed
  to the gold surfaces are hydrogen atom and sulfur atom
  respectively. The distance between the ending hydrogen atom and gold
contact is 2.15 \AA.\label{fig4}}
\end{figure}

\begin{figure}
\includegraphics[scale=0.45,bb=19 19 557 773]{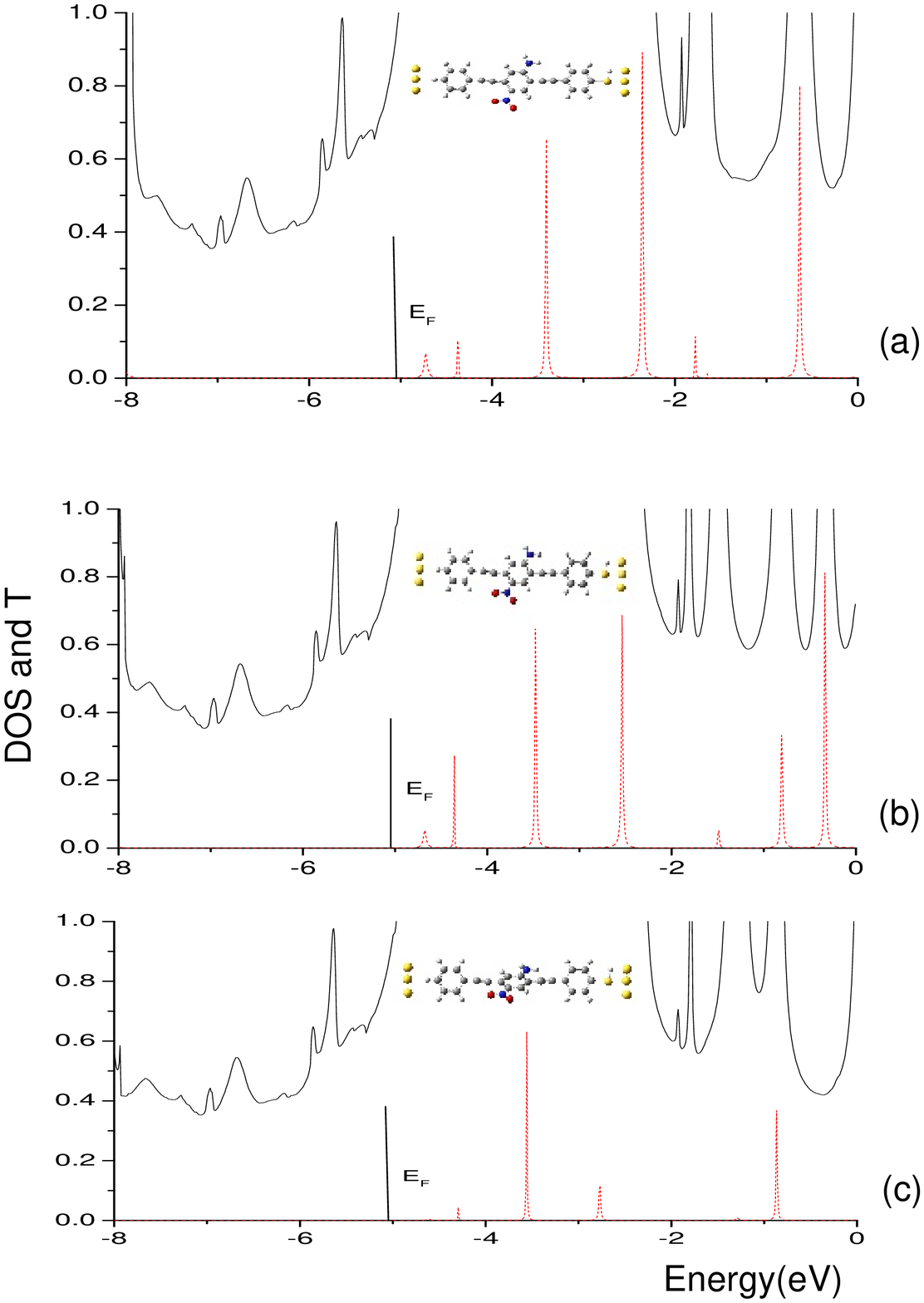}
\caption{DOS (solid) and T (dashed) as functions of energy of
$2^{'}$-amino-$4$-ethynylphenyl-$4'$-ethynylphenyl-$5'$-nitro-
$1$-benzenethiol with gold contacts in equilibrium. The atoms that
chemisorbed to the gold surfaces are hydrogen atom and sulfur atom
respectively. The distance between the ending hydrogen atom and
gold contact is 2.15 \AA. The vertical line denotes the position
of Fermi level. The rotation angle $\alpha=0^{\circ}$ (a),
$30^{\circ}$ (b), and $60^{\circ}$ (c).\label{fig5}}
\end{figure}

\begin{figure}
\includegraphics[scale=0.30,angle=-90,bb=40 19 567 805]{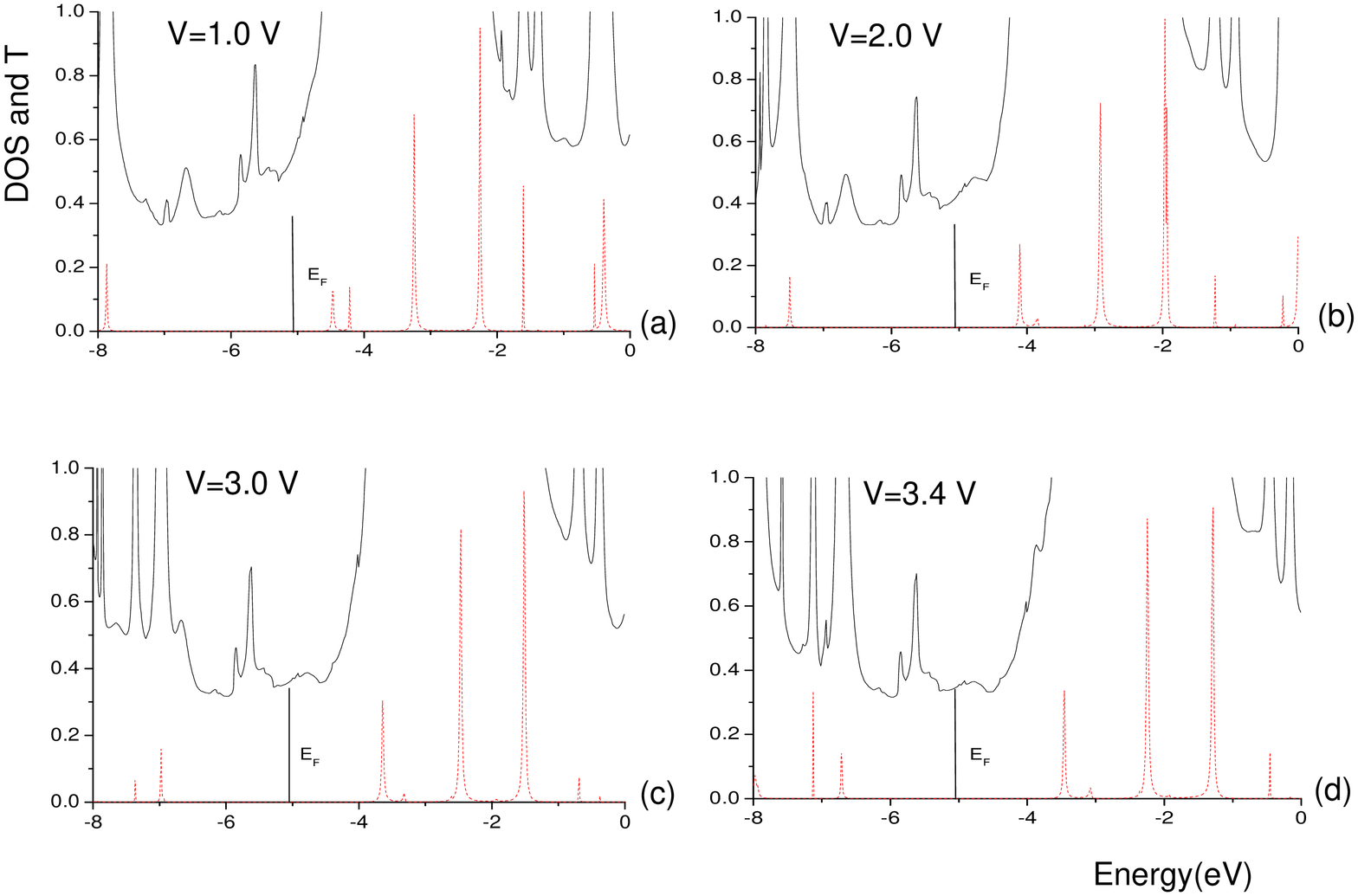}
\caption{DOS (solid) and T (dashed) as functions of energy of
$2^{'}$-amino-$4$-ethynylphenyl-$4'$-ethynylphenyl-$5'$-nitro-
$1$-benzenethiol with gold contacts corresponding to $0^{\circ}$
rotation angle with positive bias added. The atoms that
chemisorbed to the gold surfaces are hydrogen atom and sulfur atom
respectively. The distance between the hydrogen atom and gold
contact is 2.15 \AA. The vertical line denotes the position of
Fermi level. The added positive bias $1.0$ V (a), $2.0$ V (b),
$3.0$ V (c) and $3.4$ V (d).\label{fig6}}
\end{figure}

In summary, we use the self-consistent method based on the DFT and
the non-equilibrium Green's function to simulate molecular
transport. We solve the transport problem from the
first-principles theory in order to predict the transport
characteristics of some molecular devices and to identify the
experimental results from the theoretical point of view. In terms
of Gaussian03, the electronic structures for the molecular device
and the macroscopic leads are calculated on an equal footing, at
the same time, the self-consistent iteration cycle is extended
from the local molecule to the open system, by inserting the
density matrix of the open system as a subroutine. The
self-consistent iteration is run until the density matrix
converges within an acceptable accuracy. We investigate the
molecular switch reported in the experiment\cite{2}, and prove the
distinct switch function, theoretically. With one sulfur atom
replaced by the hydrogen atom, the switch effect exists in both
the low negative bias range and the high positive bias range. The
calculation result shows that the asymmetry of I-V curves can be
used to make the rectifier. Further study of the influence of gate
voltage on molecular transport characteristics is in process.

\begin{acknowledgments}
This work is supported by the National Science Foundation of China
(NSFC) under Project 90206031 and 10574024, and Special
Coordination Funds of the Ministry of Education, Culture, Sports,
Science and Technology of the Japanese Government. The author
wants to thank Ph. D candidate Y.Y. Liang for his helpful
suggestions. The author would also like to express their sincere
thanks to the support from the staff at the Center for
Computational Materials Science of IMR Tohoku University for the
use of the SR8000 G/64 supercomputer facilities.
\end{acknowledgments}

\end{document}